\pdfoutput=1

\documentclass[11pt,twoside,a4paper,cmspaper,final,collab]{cms-tdr}
\def\svnVersion{388637}\def\cmsCernNoTag{CERN-EP/2016-273}\def\cmsCernDate{\today}\def\cmsMessage{Published in the Journal of High Energy Physics as \href{http://dx.doi.org/10.1007/JHEP02(2017)048}{\doi{10.1007/JHEP02(2017)048.}}}

\begin{document}\cmsNoteHeader{EXO-16-008}

\hyphenation{had-ron-i-za-tion}
\hyphenation{cal-or-i-me-ter}
\hyphenation{de-vices}
\RCS$Revision: 381693 $
\RCS$x: svn+ssh://svn.cern.ch/reps/tdr2/notes/EXO-12-046/trunk/EXO-12-046.tex $
\RCS$Id: EXO-16-008.tex 381693 2017-01-19 21:29:47Z alverson $
\newlength\cmsFigWidth
\ifthenelse{\boolean{cms@external}}{\setlength\cmsFigWidth{0.85\columnwidth}}{\setlength\cmsFigWidth{0.4\textwidth}}
\ifthenelse{\boolean{cms@external}}{\providecommand{\cmsLeft}{top}}{\providecommand{\cmsLeft}{left}}
\ifthenelse{\boolean{cms@external}}{\providecommand{\cmsRight}{bottom}}{\providecommand{\cmsRight}{right}}

\providecommand{\NA}{\ensuremath{\text{---}}\xspace}
\providecommand{\tauh}{\ensuremath{\tau_{\mathrm{h}}}\xspace}
\newcommand{\emu}{\ensuremath{\tau_\Pe\tau_\mu}\xspace}
\newcommand{\etau}{\ensuremath{\tau_\Pe\tau_{\mathrm{h}}}\xspace}
\newcommand{\mutau}{\ensuremath{\tau_\mu\tau_{\mathrm{h}}}\xspace}
\newcommand{\ditauhad}{\ensuremath{\tauh\tauh}\xspace}
\newcommand{\taue}{\ensuremath{\tau_\Pe}\xspace}
\newcommand{\taumu}{\ensuremath{\tau_\mu}\xspace}
\newcommand{\PZprrimeSSM}{\ensuremath{\PZpr_{\mathrm{SSM}}}\xspace}
\newcommand{\PZprrimeTAT}{\ensuremath{\PZpr_{\mathrm{TAT}}}\xspace}
\newcommand{\meff}{\ensuremath{m(\tau_1,\tau_2,\ptvecmiss)}\xspace}
\newcommand{\Wjets}{{\PW}+jets\xspace}

\cmsNoteHeader{EXO-16-008}
\title{Search for heavy resonances decaying to tau lepton pairs
  in proton-proton collisions at $\sqrt{s} = 13$\TeV}

\date{\today}

\abstract{A search for heavy resonances that decay to tau
  lepton pairs is performed using proton-proton collisions at
  $\sqrt{s} = 13$\TeV.  The data were collected with the CMS detector
  at the CERN LHC and correspond to an integrated luminosity of 2.2\fbinv.  The observations are in agreement with standard
  model predictions.  An upper limit at 95\% confidence level on the
  product of the production cross section and branching fraction
  into tau lepton pairs is calculated as a function of the
  resonance mass.  For the sequential standard model, the presence of
 \PZpr bosons decaying into tau lepton pairs is excluded for \PZpr
  masses below 2.1\TeV, extending previous limits for this final state. For the topcolor-assisted technicolor model, which predicts \PZpr bosons that preferentially couple to third-generation fermions, \PZpr masses below 1.7\TeV are
  excluded, representing the most stringent limit to date.}

\hypersetup{%
pdfauthor={CMS Collaboration},%
pdftitle={Search for heavy resonances decaying to tau lepton pairs  in proton-proton collisions at sqrt(s) = 13 TeV},%
pdfsubject={CMS},%
pdfkeywords={CMS, exotica, tau, zprime, ADD}
}

\maketitle
\section{Introduction}
\label{Introduction}
The standard model (SM) of particle physics is a successful theory
that can explain many experimental observations involving weak,
electromagnetic, and strong interactions. Nevertheless, it cannot be
an ultimate theory of nature since it fails, for instance,
to provide an explanation for the mass of neutrinos and lacks a particle candidate for dark matter.
Many models of physics beyond the SM have therefore been proposed.
Such models often predict new heavy particles that could be observed at the CERN LHC.

A straightforward way to extend the SM gauge structure is to include an
additional U(1) group with an associated neutral gauge
boson, denoted \PZpr.  The universality of couplings is not necessary
for new gauge bosons.  Indeed, models exist that incorporate
generation-dependent couplings, resulting in \PZpr bosons that
preferentially decay into fermions of the third
generation~\cite{langacker, Lynch}.  Such models motivate a search for \PZpr
resonances that decay to a pair of $\tau$ leptons.

In particular, extensions to the SM proposed as an explanation for the high mass of the top quark predict \PZpr bosons that typically couple to
third-generation fermions~\cite{Lynch}. Examples are the topcolor-assisted technicolor (TAT) models~\cite{TopcolorAssistedTechnicolor1,TopcolorAssistedTechnicolor2}.
A widely used benchmark model in searches for \PZpr bosons is the sequential
standard model (SSM)~\cite{Altarelli}, which predicts a neutral spin-1 \PZpr boson,
denoted \PZprrimeSSM, with the same couplings to quarks and leptons as
the SM Z boson.

Results of direct searches for heavy $\tau\tau$ resonances in
proton-proton (pp) collisions at either $\sqrt{s}=7$ or 8\TeV have been
reported by the ATLAS and CMS Collaborations and exclude \PZprrimeSSM
masses below 2.0\TeV~\cite{ZpCms7,ZpAtlas7, ZpAtlas8}.  The most
stringent mass limits on \PZprrimeSSM production, set by ATLAS and CMS
in searches for a narrow resonance decaying into an
$\Pep\Pem$ or $\Pgmp\Pgmm$ pair, are 3.4~\cite{Aaboud:2016cth}
and 3.2~\cite{ZpCMSmumuee13} \TeV, respectively.

In this paper we report on a search for physics beyond the SM in
events containing a pair of high transverse momentum (\pt) oppositely
charged $\tau$ leptons.
Four $\tau\tau$ final states, $\tau_\Pe \tau_{\mu}$, $\tau_\Pe \tauh$,
$\tau_{\mu} \tauh$, and \ditauhad, are selected, where $\tau_\ell$ ($\ell = \Pe,\mu$) and \tauh
refer to the leptonic and hadronic decay modes of the $\tau$ lepton, respectively.
The study is based on a data sample of
pp collisions at $\sqrt{s} = 13\TeV$ recorded with the CMS detector at the LHC.
The sample corresponds to an integrated luminosity of 2.2\fbinv.

\section{The CMS experiment}
\label{The CMS experiment}
The central feature of the CMS apparatus is a superconducting
solenoid, of 6\unit{m} internal diameter, providing a field of 3.8\unit{T}. Within
the field volume are the inner tracker, the crystal electromagnetic
calorimeter (ECAL), and the brass and scintillator hadron calorimeter (HCAL).
The inner tracker is composed of a pixel detector and a silicon strip
tracker, and measures charged-particle trajectories in the
pseudorapidity range $\abs{\eta} < 2.5$.
The finely segmented ECAL
consists of nearly 76 000 lead-tungstate crystals that provide
coverage up to $\abs{\eta} = 3.0$.
The HCAL consists of a sampling calorimeter, which utilizes alternating layers of
brass as an absorber and plastic scintillator as an active material,
covering the range $\abs{\eta}<3$, and is extended to $\abs{\eta}<5$ by a forward hadron
calorimeter.
The muon system covers $\abs{\eta} < 2.4$
and consists of four stations of gas-ionization muon detectors
installed outside the solenoid and sandwiched between the layers of
the steel return yoke.  A detailed description of the CMS detector,
together with a definition of the coordinate system and relevant
kinematic variables, is given elsewhere~\cite{CMS}.

Events of interest are selected using a two-tiered trigger system~\cite{Khachatryan:2016bia}.
The first level (L1), composed of custom hardware processors, uses information from the calorimeters and muon detectors
to select events at a rate of around 100\unit{kHz} within a time interval of less than 4\mus.
The second level, known as the high-level trigger (HLT), consists of a farm of processors running a version of
the full event reconstruction software optimized for fast processing, and reduces the event rate to
less than 1\unit{kHz} before data storage.

\section{Object reconstruction and identification}
\label{Object reconstruction and identification}
A particle-flow (PF) algorithm~\cite{PFTausReco,PFT-10-001} is used to combine
information from all CMS subdetectors in order to reconstruct and
identify individual particles in the event: muons, electrons, photons,
and charged and neutral hadrons.  The resulting set of particles is
used to reconstruct the \tauh candidates, jets, missing transverse momentum,
and the isolation variables described below.  The primary vertex
of an event is chosen to be the reconstructed vertex with the largest $\pt^2$ sum
of associated tracks.

The $\tau_\Pe$ and $\tau_{\mu}$ are reconstructed and identified with the
usual techniques for muons and electrons.
Electrons are reconstructed as energy deposits in the ECAL associated
with tracks in the tracking detector~\cite{electron8TeV,ZpCMSmumuee13}.
Requirements on energy deposits in the calorimeter and on the number
of hits in the inner tracker are imposed to distinguish
electrons produced in hard scattering processes from charged pions and
from electrons produced through photon conversions.

Muons are reconstructed using the inner tracker and the muon
detectors~\cite{muonRun1,muon13TeV}. Quality requirements, based on the minimum
number of hits in the inner tracker and
muon detectors, are applied to suppress backgrounds from the
decays-in-flight of light-flavor hadrons, and from hadron shower remnants
that reach the muon system.

Both the electron and muon selections impose an isolation requirement
to suppress both jets erroneously identified as leptons and genuine leptons
from hadron decays. The isolation criterion for light leptons is based
on the variable $I_{\ell}$, which is the scalar \pt sum, divided by the lepton \pt, of charged
and neutral PF candidates within a cone of radius $\DR =
\sqrt{\smash[b]{(\Delta\eta)^2 + (\Delta\phi)^2}} = 0.3$ around the lepton
direction at the interaction vertex, where
$\phi$ is the azimuthal angle.  The sum excludes the lepton under
consideration as well as charged particles from additional pp
interactions within the same or a nearby bunch crossing (``pileup'').
The contribution of neutral particles from pileup is accounted for by
subtracting from the isolation sum a term given by the scalar \pt sum
of charged hadrons from pileup vertices that appear within the
isolation cone, multiplied by a factor of 0.5 to account for the ratio
of neutral to charged hadron production.  The isolation criterion is
$I_{\ell}<0.15$.

The \tauh candidates are reconstructed using the hadrons-plus-strips
algorithm~\cite{TauAt8TeV, CMS:2016gvn}, which is designed to
optimize the performance of \tauh reconstruction and identification by
considering specific $\tau$ lepton decay modes.
Individual hadronic
$\tau$ decay modes are reconstructed separately. The signatures distinguished by the algorithm
are: a single charged hadron, a charged hadron plus up to two neutral pions, and three charged hadrons.
Reconstructed \tauh
leptons are required to be isolated to reduce background from
misidentified light-quark or gluon jets.  The isolation criterion for
\tauh leptons is based on the scalar \pt sum $S_\tau$ of charged
and neutral PF candidates within a cone of radius $\Delta R = 0.5$ around
the \tauh direction, excluding the \tauh candidate and charged tracks
from pileup.  The contribution of neutral particles from pileup is
accounted for by subtracting from $S_\tau$ the scalar \pt sum of
charged particles from pileup interactions that appear within a cone
of radius $\Delta R=0.8$ around the \tauh candidate, multiplied by a
factor of 0.2~\cite{CMS:2016gvn}.  The factor of 0.2 is chosen to render the \tauh
identification efficiency insensitive to the level of pileup.  The
isolation criterion is $S_\tau<0.8\GeV$. The \tauh candidates are further distinguished from electrons and muons using dedicated
algorithms, referred to as discriminators.  The algorithm to
discriminate a \tauh lepton from an electron utilizes observables that
quantify the compactness and shape of energy deposits in the ECAL, to
distinguish electromagnetic from hadronic showers, in combination with
observables that are sensitive to the amount of bremsstrahlung
radiation emitted along the highest $\pt$ track of the \tauh candidate, and observables that are
sensitive to the overall particle multiplicity. The discriminator
against muons requires that no hits in the muon system be matched to the \tauh candidate.
For a \tauh with $\pt >20\GeV$ and $\abs{\eta} < 2.1$, the identification efficiency is approximately 55\%. The
probability for a light-quark or gluon jet, electron, and muon to be misidentified as a \tauh is approximately 1\%, 0.2\%, and 0.03\%.

The jets are reconstructed using the anti-\kt jet
algorithm~\cite{Cacciari:2008gp,Cacciari:2011ma} with a distance
parameter of 0.4.  In order to tag jets from b quark decays, the
combined secondary vertex (CSVv2) algorithm at the loose working point
is used~\cite{Btag8,Btag13}. This algorithm is based on the reconstruction
of secondary vertices, together with track-based lifetime information.
For b quark jets with $\pt >30\GeV$ and $\abs{\eta} < 2.4$, the
identification efficiency is approximately 85\%, while the probability for a
light-quark or gluon (charm quark) jet to be misidentified as a b quark jet is
approximately 10\% (20\%).

The missing transverse momentum vector \ptvecmiss is defined as the
negative of the vector sum of transverse momenta of all PF candidates
reconstructed in the event. The magnitude of this vector is referred
to as \MET.  The raw \MET value is modified to account for corrections
to the energy scale of all the reconstructed jets in the
event~\cite{MET}.
Events that contain large values of \MET as a consequence of instrumental effects,
such as calorimeter noise, beam halo, or jets
near nonfunctioning channels in the calorimeters, are removed from
the analysis.

\section{Signal and background Monte Carlo samples}
\label{Signal and background MC samples}

The most important sources of background arise from the production of
Drell--Yan events ($\Z/\gamma^{*} \to \tau^+\tau^-$), {\PW}
bosons in association with one or more jets (\Wjets), top quark pairs
(\ttbar), quantum chromodynamics (QCD) multijet events, and diboson events (\PW\PW, \PW\cPZ, \cPZ\cPZ).
Although the Drell--Yan
background peaks around the Z pole mass, its tail extends into the
high-mass region where a signal might be present. The \Wjets events are
characterized by an isolated lepton from the decay of the W boson and
an uncorrelated jet misidentified as a light lepton or \tauh
lepton. Background from \ttbar events contains one or
two b quark jets, in addition to isolated $\tau_\ell$ or \tauh candidates.
Background from diboson events produces isolated leptons if the gauge
bosons decay leptonically.  Should the gauge bosons decay hadronically, one of the jets may be misidentified as a \tauh lepton.
Finally, QCD multijet background is characterized by jets with
low charged-track multiplicity that can be misidentified as
$\tau_\ell$ or \tauh candidates.

Monte Carlo (MC) event generators are used to simulate the signal and
SM backgrounds.  The Drell--Yan, \Wjets, and \ttbar processes are
generated with the \MADGRAPH{5}\_a\MCATNLO program~\cite{Madgraph}.  The
\PYTHIA~8.2 program is used to generate the diboson and signal events~\cite{Sjostrand:2014zea}.  The signal events are generated
with \PZpr masses ranging from 0.5 to 3.0\TeV, with a step size of
0.5\TeV.
The expected signal yields are rescaled to
next-to-leading order (NLO) accuracy using a $K$-factor of 1.3~\cite{ZpCMSmumuee13}. The electroweak NLO corrections are small in the range of masses considered. The SM
samples are normalized using the most accurate cross section
calculations currently
available~\cite{Alioli:2009je,Re:2010bp,Madgraph,Melia:2011tj,Beneke:2011mq,
  Cacciari:2011hy,Baernreuther:2012ws,Czakon:2012zr,Czakon:2012pz,Czakon:2013goa,
  Gavin:2012sy,Gavin:2010az}, generally with NLO or next-to-next-leading order (NNLO)
accuracy.  The NNPDF~3.0~\cite{Ball:2014uwa} parton distribution
functions (PDF) are used, and all simulated samples use the {\PYTHIA}
program with the CUETP8M1 tune~\cite{Tunes} to describe parton showering and hadronization.  The
simulation of pileup is performed by superimposing minimum
bias interactions onto the hard scattering process, matching the
pileup profile in data. The mean number of interactions in a single bunch crossing in the analysed
data set is approximately 14. The MC-generated events are propagated
through a {\GEANTfour}-based simulation~\cite{Geant4} of the CMS
apparatus.

\section{Event selection}
\label{Event selection}
The requirements described below define the signal region.
A new heavy neutral gauge boson decaying into a $\tau$ lepton pair
would be characterized by an excess above the SM expectation for the
rate of events with two high-\pt, oppositely charged, isolated $\tau$
lepton candidates.  Single-lepton triggers are used to select
$\tau_\Pe \tau_{\mu}$, $\tau_\Pe \tauh$, and $\tau_{\mu} \tauh$
events, while a trigger requiring at least two \tauh candidates
at the L1 and HLT levels is
used to select \ditauhad events. The triggers are designed to allow the use of
the background estimation methods outlined in Section~\ref{Background estimation}.
Electrons (muons) are required to
have $\pt > 35\,(30)\GeV$. The \tauh candidates are required to have
$\pt > 20$ and $60\GeV$ in the $\tau_{\ell} \tauh$ and
\ditauhad channels, respectively.
The $\tau$ lepton candidate \pt is defined by the vector \pt sum of its visible decay products.
Both $\tau_{\ell}$ and \tauh
candidates must have $\abs{\eta} < 2.1$ and satisfy isolation requirements
to mitigate background from misidentified jets.  The \pt thresholds on
the $\tau_\Pe$, $\tau_{\mu}$, and \tauh candidates are chosen such
that the trigger efficiency is about 90\% or higher in each channel
considered.

The $\tau\tau$ pairs are formed from oppositely charged candidates with
$\DR(\tau_1,\tau_2) > 0.5$. The \tauh charge is reconstructed from
the sum of the charges of the associated tracks used to reconstruct the decay mode
and is required to be $\pm$1. Owing to the large invariant mass of the
$\tau\tau$ resonances assumed for this study, the two $\tau$ candidates are
expected to be back-to-back. Events are therefore required to satisfy
$\cos{\Delta\phi(\tau_1,\tau_2)} < -0.95$.
An additional requirement of $\MET >30\GeV$ is applied to preferentially select events with neutrinos
from the $\tau$ lepton decays rather than apparent $\MET$ due to mismeasurement of jet~\pt. The signal efficiency of this
requirement is 85\% efficiency or more, depending on the $\PZpr$ boson mass.

The direction of \ptvecmiss is required to be consistent with the
expectation for a pair of high-\pt $\tau$ lepton decays, to reduce the
background from events with {\PW} bosons (primarily \Wjets and \ttbar events).  This requirement is
implemented through a variable known as ``CDF-$\zeta$"~\cite{cdfzeta},
referred to below as the $\zeta$ variable.  This variable is defined
by considering a unit vector, denoted the $\hat{\zeta}$ axis, along
the bisector between the \pt directions of the two $\tau$ lepton
candidates.
Two projection variables for the visible $\tau$ lepton decay products and \ptvecmiss are then
constructed: $p_{\zeta}^{\text{vis.}} = (\ptvec^{\tau_1} +
\ptvec^{\tau_2}) \cdot \hat{\zeta}$ and $p_{\zeta} = (\ptvec^{\tau_1}
+ \ptvec^{\tau_2} + \ptvecmiss) \cdot \hat{\zeta}$. In contrast to signal events in which
$p_{\zeta}^{\text{vis.}}$ and $p_{\zeta}$ are strongly correlated, these two variables are nearly
independent in events with a W boson because in this case the direction and
magnitude of \ptvecmiss are
correlated with those of the lepton, but not with those of the jet.
Events are selected by requiring $\zeta = p_{\zeta} -3.1
p_{\zeta}^{\text{vis.}} > -50\GeV$.  Residual contributions from
\ttbar events are reduced by selecting events without a tagged b jet.

To distinguish more effectively between signal and background events,
the visible $\tau$ lepton energies and momenta $E_{\tau_i}$ and
$\vec{p}_{\tau_i}$ (with $i=1,2$), along with \ptvecmiss, are used
to reconstruct a mass value $m$, defined as
\begin{equation}
  \meff = \sqrt{(E_{\tau_1}+E_{\tau_2}+\MET)^{2}-
    (\vec{p}_{\tau_1}+\vec{p}_{\tau_2}+\ptvecmiss)^{2}}.
\label{eq:ditaumass}
\end{equation}
We do not apply a selection requirement on the mass variable of
Eq.~(\ref{eq:ditaumass}), but instead utilize it to search for a broad
enhancement in the \meff distribution consistent with new physics.

The product of signal acceptance and efficiency for $\PZpr \to
\tau\tau$ events varies with the $\PZpr$ boson mass. The $\mutau$ and \ditauhad channels provide the largest product of acceptance and efficiency, while the $\tau_\Pe \tau_{\mu}$ channel has the lowest.
For the combination of all four
channels the product of acceptance and efficiency amounts to 6.3, 13, and 14\%, respectively, for
$m\left(\PZprrimeSSM\right)$ = 0.5, 1.5, and 3.0\TeV.

\section{Background estimation}
\label{Background estimation}
To estimate the background contributions in the signal region,
techniques based on data control regions are employed wherever
possible.  The strategy when using such a technique is to modify the
standard event selection requirements in order to define samples
enriched with events from specific sources of background. These control regions are used to
model the \meff shape of the backgrounds and to measure the
probability for an event to satisfy the selection requirements.
Contributions from additional background sources in a
given control region are subtracted using their predictions from simulation. The background
estimation methods based on data control regions are validated by
determining the ability of the method, applied to simulated samples,
to predict correctly the true number of background events and the
shape of the \meff distribution. In some cases, for backgrounds with misidentified leptons,
the background estimation methods are also validated with the data.
In such tests, the agreement between the relevant quantities is always within the statistical uncertainties, and is at the level of 20\% or better, depending on the channel.
In cases where an approach based on a data
control region is not possible, or for backgrounds with small expected
contributions, we rely on simulation. For the most relevant SM
contributions evaluated from simulation, we verify that the MC
prediction is in agreement with the data in background-enhanced
regions.

The QCD multijet background is relevant for both the \ditauhad and
$\tau_{\ell}\tauh$ channels, where it represents more than 80\% or 20\% of
the total background, respectively. The contribution from QCD multijet
events in the $\tau_\Pe \tau_{\mu}$ channel is approximately 5\% ($<1$\%) for
$\meff > 85\,(300)\GeV$.
In the \ditauhad final state,
this background is evaluated from the like-sign $\tau\tau$ mass
distribution ($> 98$\% purity of QCD multijet events), which is scaled using the opposite-sign-to-like-sign
ratio measured in a control region where the \MET requirement is
inverted ($\MET < 30\GeV$).  In the $\taue\tauh$ final state, the QCD
multijet background is evaluated using the mass shape reconstructed
from a data sample with a nonisolated \tauh. This mass shape is weighted by the
probability for a jet to satisfy the \tauh isolation criterion, which
is measured from a sample of like-sign $\tau\tau$ candidates.
The systematic uncertainties in these estimates are discussed in Section~\ref{Systematic uncertainties}.

Background from \Wjets events is important in the $\tau_{\ell} \tauh$
channels, representing about 40 (45)\% of the total background when $\ell$ is an electron (muon).
The contribution from \Wjets events in the $\tau_\Pe \tau_{\mu}$ and \ditauhad channels is $<$1\% for $\meff > 300\GeV$.
To estimate this background, we take the mass distribution of events
selected in data with one nonisolated \tauh, subtract the QCD multijet
background as determined from a data control region, subtract other
backgrounds as determined from simulation, and weight the distribution
by the probability for a jet to satisfy the \tauh isolation
criterion. This probability is measured in a data control region for
which the $\zeta$ and $\cos{\Delta\phi(\tau_1,\tau_2)}$ requirements
are inverted ($\zeta < -50\GeV$ or $\cos{\Delta\phi} > -0.95$). The
sample used to determine the QCD multijet contribution to the \Wjets
control region contains a small expected contribution from \Wjets
events, which is subtracted using simulation.
The systematic uncertainties in these estimates are discussed in Section~\ref{Systematic uncertainties}.

The \ttbar background is relevant for the $\tau_\Pe \tau_{\mu}$
channel, where it represents more than 30\% (70\%) of the total
background for $\meff > 85$ (300)$\GeV$. Its contribution to the other
channels is $<$2\%\,($<1$5\%) for $\meff > 85\,(300)\GeV$. High-purity samples of \ttbar events are obtained by requiring the presence
of at least one b-tagged jet with $\pt > 20\GeV$. The
\ttbar prediction from simulation agrees with the observed yield and
mass shape in the control sample, with a statistical uncertainty of
$\sim$8\% in the ratio of the simulated to the observed yield. Thus the \ttbar
prediction in the signal region is based on simulation with an
additional systematic uncertainty of 8\%.

The SM Drell--Yan background contributes to all final states, ranging
from about 10\% of the total background in the \ditauhad channel to
40\% in the other channels. It is estimated from simulation, after
comparing the expectations from simulation with data in low-mass
control regions where no signal is expected~\cite{ZpCms7}, specifically regions where $\MET<30\GeV$, and
either $m(\tau_1,\tau_2) < 100\GeV$ (\ditauhad channel) or
$\meff < 200\GeV$ (the other channels).
The data and simulation are found
to be in agreement within the systematic uncertainties discussed in
Section~\ref{Systematic uncertainties}.

The remaining backgrounds described in Section~\ref{Signal and
  background MC samples} are estimated using simulation.

\section{Systematic uncertainties}
\label{Systematic uncertainties}
The main source of systematic uncertainty is the
uncertainty in the estimation of the background, due to the limited
number of events in the data control regions. This results in
uncertainties
ranging from 8\% for the contribution from dominant backgrounds such as that from \Wjets events in the $\tau_{\ell}\tauh$
channels, to 68\% for the small contribution of QCD multijet
events in the $\tau_{\mu} \tauh$ channel.  Nondominant backgrounds in the
control regions, including the resulting uncertainties in the control
sample purities, make a minor contribution to the overall systematic
uncertainty.

The efficiencies for electron and muon reconstruction, identification,
and triggering have an uncertainty of approximately 2\%,
measured using $\cPZ\to\ell^+\ell^-$
events~\cite{electron8TeV, muon13TeV}.
For the uncertainty in the
description by the simulation of the identification
efficiency for high-\pt electrons (muons), an uncertainty
of 6\%\,(7\%), independent of \pt, is assigned, as described
in Ref.~\cite{ZpCMSmumuee13}.
The systematic uncertainty on \tauh trigger efficiency is 5\% per \tauh candidate, as measured with $\Z\to\mutau$ events triggered by
single-muon triggers.
Systematic effects associated with \tauh
identification are measured using the visible $\tau\tau$ mass distribution
around the \Z boson peak as well as off-mass-shell virtual {\PW} bosons selected
with a single \tauh candidate and large $\MET$~\cite{CMS:2016gvn}.
The resulting uncertainty is 6\% per \tauh
candidate.  An additional systematic uncertainty, which dominates for
high-\pt \tauh candidates, is related to the
confidence that the MC simulation correctly models the identification
efficiency, and is validated
with high-\pt jet and electron candidates that produce $\tauh$-like signatures in the detector.
This additional uncertainty increases linearly with \pt and
amounts to 20\% per \tauh candidate at $\pt = 1$\TeV
(correlated between both candidates in the \ditauhad channel),
resulting in an uncertainty of 4\%\,(10\%) for a reconstructed mass of 0.5\TeV in the $\tau_{\ell} \tauh$
(\ditauhad) channel, and 12\%\,(25\%) for a mass of 2\TeV.
The uncertainty in the integrated luminosity measurement is 2.7\%~\cite{REFLUMI}.  Uncertainties
that affect the \meff mass shape include the electron, muon, and \tauh
energy scales, and the jet and \MET energy scales and resolutions.  The
uncertainty in the probability for a light quark or gluon jet to be
misidentified as a b quark jet (20\%) is also considered, and has a
$\sim$3\% effect on the signal acceptance and a negligible effect on
the \meff mass shape.  The uncertainty in the signal acceptance
associated with the PDFs is evaluated in accordance with the PDF4LHC
recommendations~\cite{Butterworth:2015oua} and amounts to a few
percent.

The systematic uncertainties in the \Wjets, Drell--Yan, QCD multijet,
and \ttbar background normalizations are approximately 9\%\,(9\%), 10\%\,(19\%), 68\%\,(20\%), and 8\%\,(8\%), respectively, in the $\tau_{\ell}
\tauh$ (\ditauhad) channel. The total systematic uncertainty in the SM
background yield ranges from $<$10\% at low mass in the $\tau_{\ell}
\tauh$ channels to $\sim$100\% at high mass in the \ditauhad
channel. The normalization uncertainties have a fully correlated
effect across all mass bins, while the shape uncertainties account for
systematic migrations of expected yields among neighboring bins in \meff.

\section{Results}
\label{Results}
\begin{figure*}[htbp!]
\centering
\includegraphics[width=0.49\textwidth]{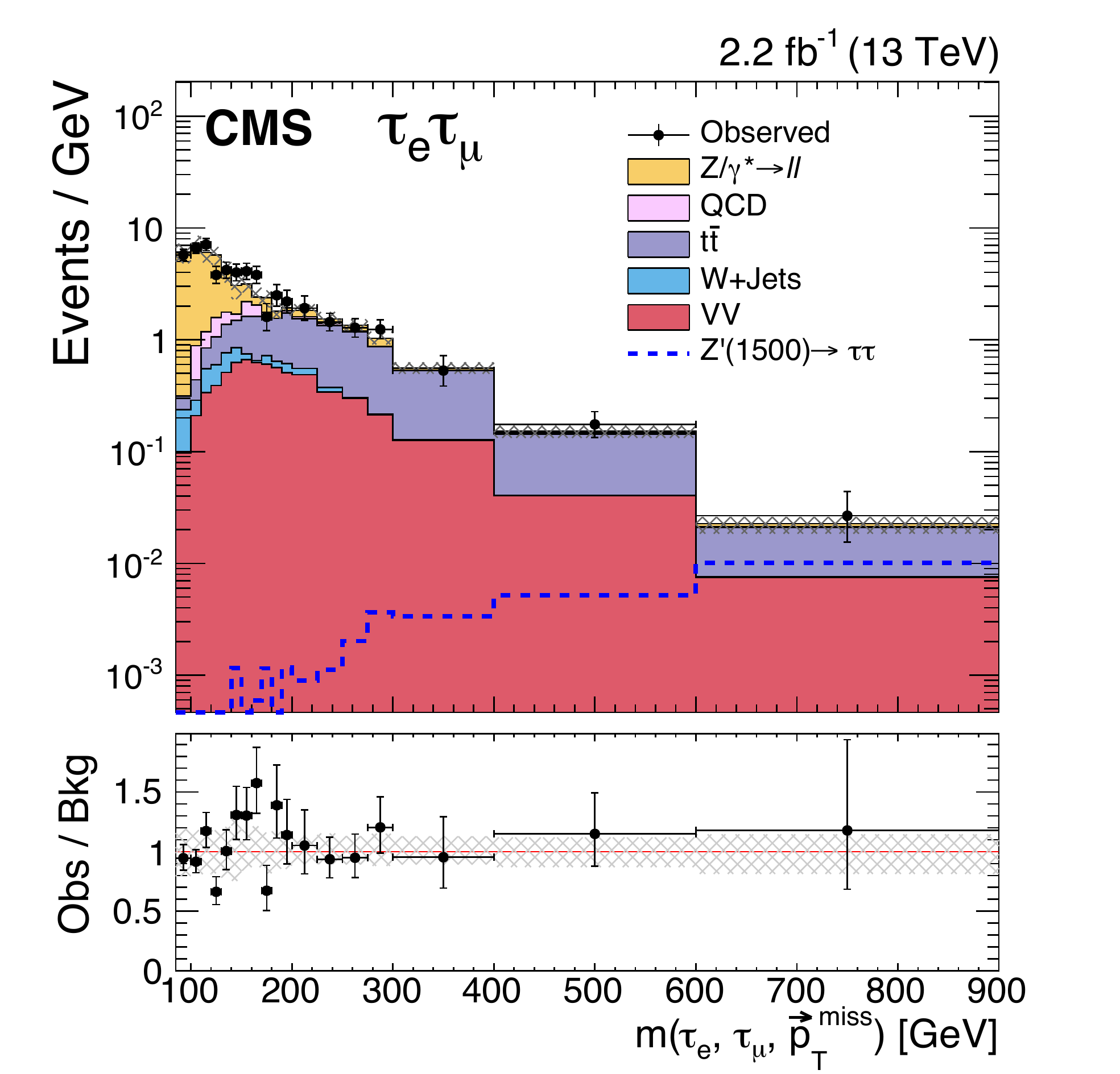}
\includegraphics[width=0.49\textwidth]{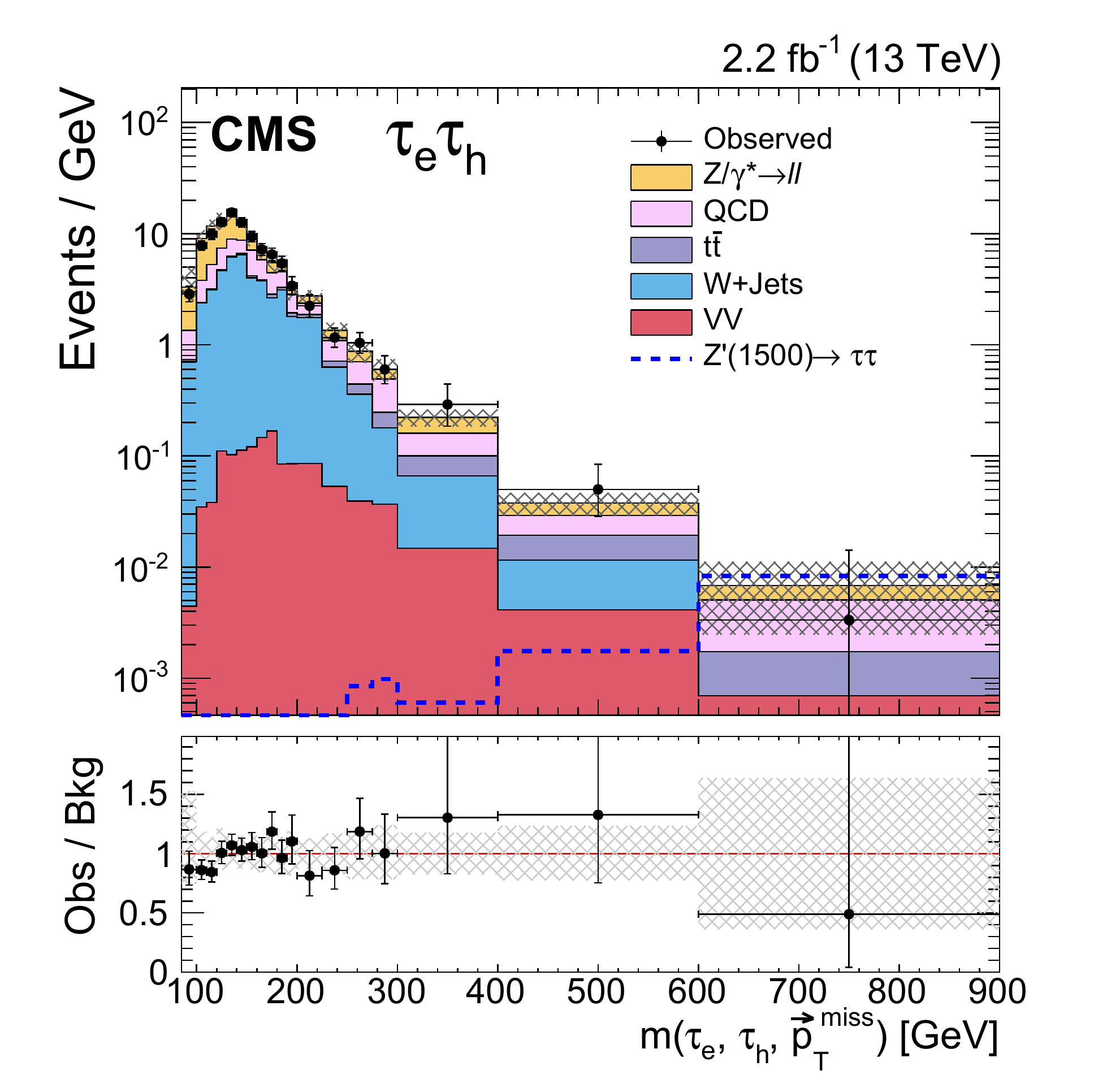} \\
\includegraphics[width=0.49\textwidth]{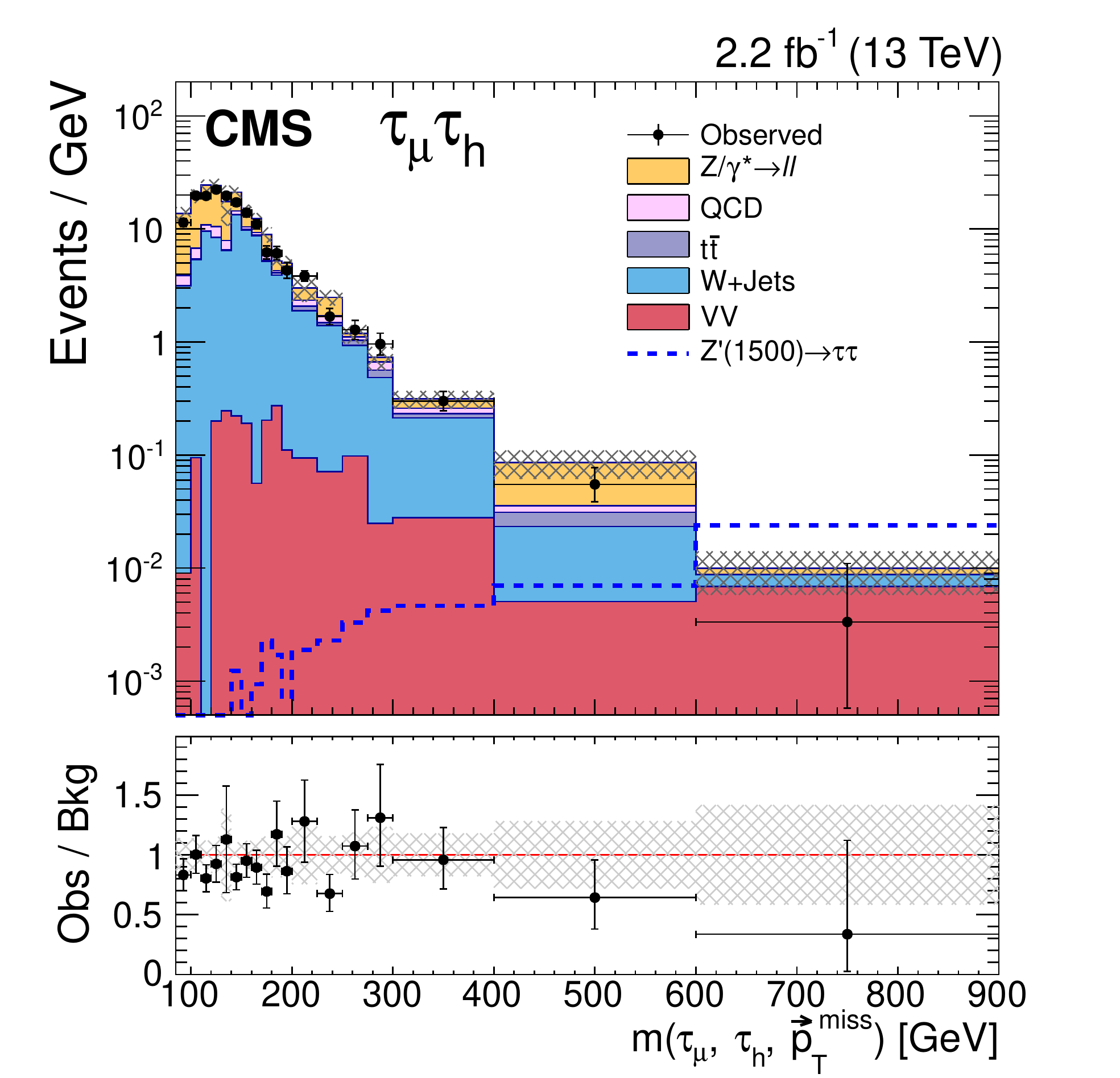}
\includegraphics[width=0.49\textwidth]{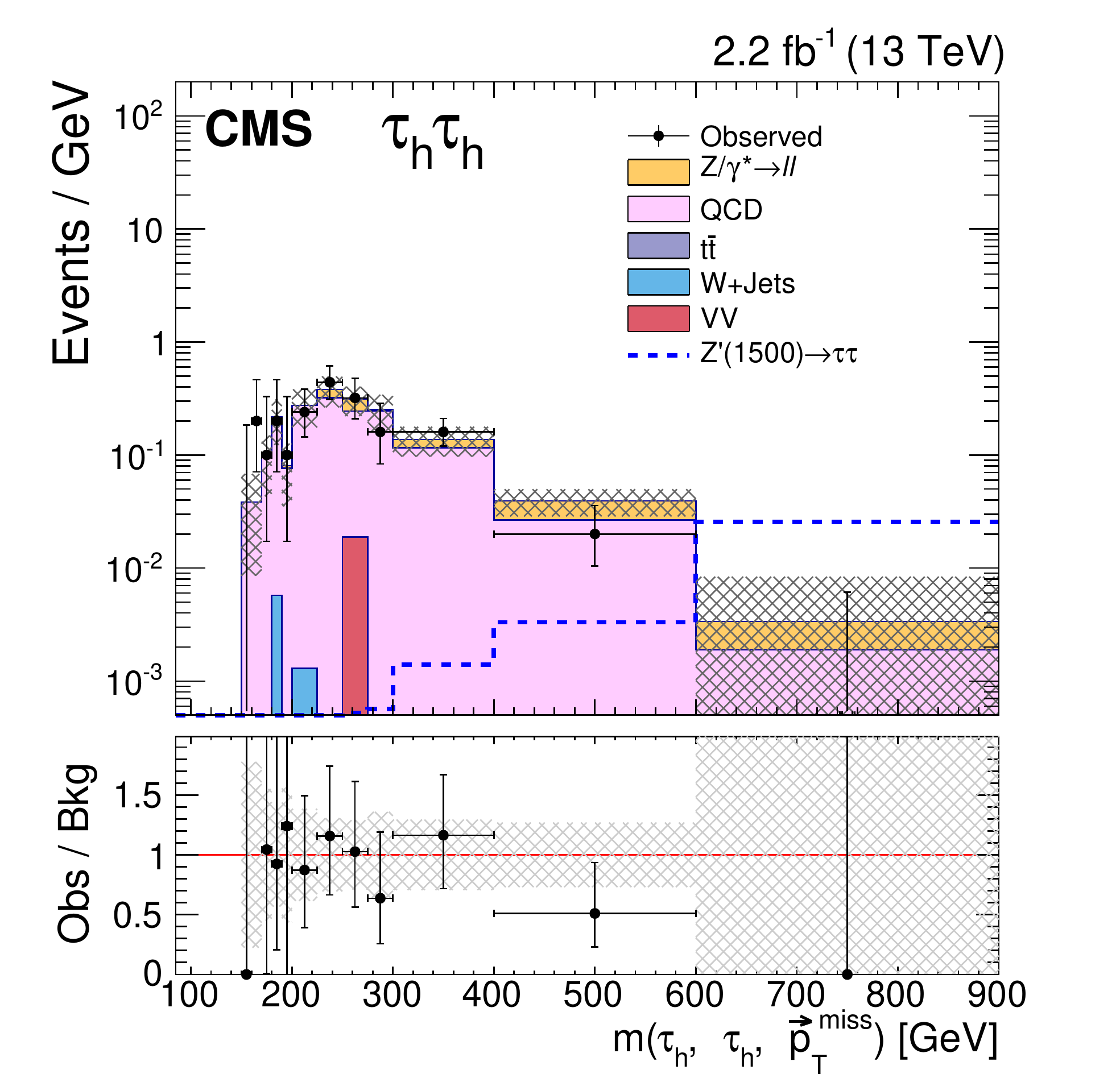}
\caption{Observed \meff distribution in the signal region compared to the expected SM
  backgrounds for the (top left) \emu, (top right) \etau, (bottom
  left) \mutau, and (bottom right) \ditauhad channels.  The dashed
  histogram shows the distribution expected for a \PZprrimeSSM boson with
  mass 1500\GeV.  The rightmost bins also include events with
  $\meff>900\GeV$, and are normalized to the displayed bin width.  The lower panel shows the ratio of the observed
  number of events to the total background prediction. The shaded bands represent the total uncertainty in the background prediction.}
\label{fig:masses}
\end{figure*}

The upper section of Table~\ref{table:EvtSR} lists the estimated
background yields and the total number of observed events for each
channel, integrated over all values of reconstructed mass.  The lower
section of Table~\ref{table:EvtSR} lists the results for $\meff >
300\GeV$, where the signal is primarily expected.  The distributions
of \meff are shown in Fig.~\ref{fig:masses} for all four channels.

\begin{table*}[htbp!]
\topcaption{Numbers of events observed in data compared to the
  expected background yields and to the predicted numbers of signal
  events for \PZprrimeSSM masses of 1.0, 1.5, and 2.0\TeV.  The
  upper section of the table presents the inclusive yields.  The lower
  section presents the yields after the requirement $\meff > 300\GeV$.  The
  uncertainties quoted in the background yields represent the combined
  statistical and systematic uncertainties.}
\centering
\newcolumntype{x}{D{,}{\,\pm\,}{3.3}}
\begin{tabular}{lxxxx}
\hline
Process               & \multicolumn{1}{c}{\emu}         & \multicolumn{1}{c}{\etau}           &\multicolumn{1}{c}{ \mutau}           &\multicolumn{1}{c}{ \ditauhad }          \\
\hline
Drell--Yan            & 321,37      & 375,40 & 882,130 & 8,3 \\
\Wjets               & 19,6	   & 456,35 & 916,96  & 0.1,0.1 \\
Diboson              & 108,11      & 18,4  & 29,7   & 0.5,0.5 \\
\ttbar               & 223,20      & 26,6  & 26,7   & \multicolumn{1}{c}{\NA}  \\
QCD multijet         & 36,16      & 250,50 & 122,84  & 49,13  \\
\hline
Total                & 707,47      & 1125,73 & 1976,180 & 58,13  \\
\hline
Observed             & \multicolumn{1}{c}{728}	           & \multicolumn{1}{c}{1113}            & \multicolumn{1}{c}{1807}             & \multicolumn{1}{c}{55} \\
\hline
\PZprrimeSSM (1.0\TeV) & 24.7,1.9 & 19.1,1.4 & 53,4 & 45,3   \\
\PZprrimeSSM (1.5\TeV) & 4.7,0.3 & 3.0,0.1 & 9.4,0.4 & 8.6,0.4   \\
\PZprrimeSSM (2.0\TeV) & 1.2,0.05 & 0.77,0.04 & 2.3,0.1 & 2.1,0.1   \\

\hline
\\
\end{tabular}
\begin{tabular}{lxxxx}
\hline
Process               & \multicolumn{1}{c}{\emu}         & \multicolumn{1}{c}{\etau}           & \multicolumn{1}{c}{\mutau}           & \multicolumn{1}{c}{\ditauhad}           \\
\hline
Drell--Yan            & 4,3 & 9,4 & 16,4 & 5,2   \\
\Wjets               & 0.2,0.5 & 7,5 & 23,9 & 0.004,0.004 \\
Diboson              & 23,5 & 3,2 & 6,3 & 0.02,0.02 \\
\ttbar               & 65,12 & 5,3 & 4,2 &\multicolumn{1}{c}{ \NA}  \\
QCD multijet         & 0.8,1.0 & 9,3 & 4,3 & 18,6 \\
\hline
Total                & 93,13 & 33,8 & 51,11 & 23,6    \\
\hline
Observed             & \multicolumn{1}{c}{96} & \multicolumn{1}{c}{40} & \multicolumn{1}{c}{42} & \multicolumn{1}{c}{20}             \\
\hline
\PZprrimeSSM (1.0\TeV) & 21.1,1.6 & 18.1,1.3 & 49,4 & 44,3   \\
\PZprrimeSSM (1.5\TeV) & 4.4,0.3 & 2.9,0.1 & 9.0,0.4 & 8.5,0.4    \\
\PZprrimeSSM (2.0\TeV) & 1.2,0.05 & 0.77,0.04 & 2.3,0.1 & 2.1,0.1   \\
\hline
\end{tabular}
\label{table:EvtSR}
\end{table*}

\begin{figure*}[htbp!]
  \centering
    \includegraphics[width=0.42\textwidth]{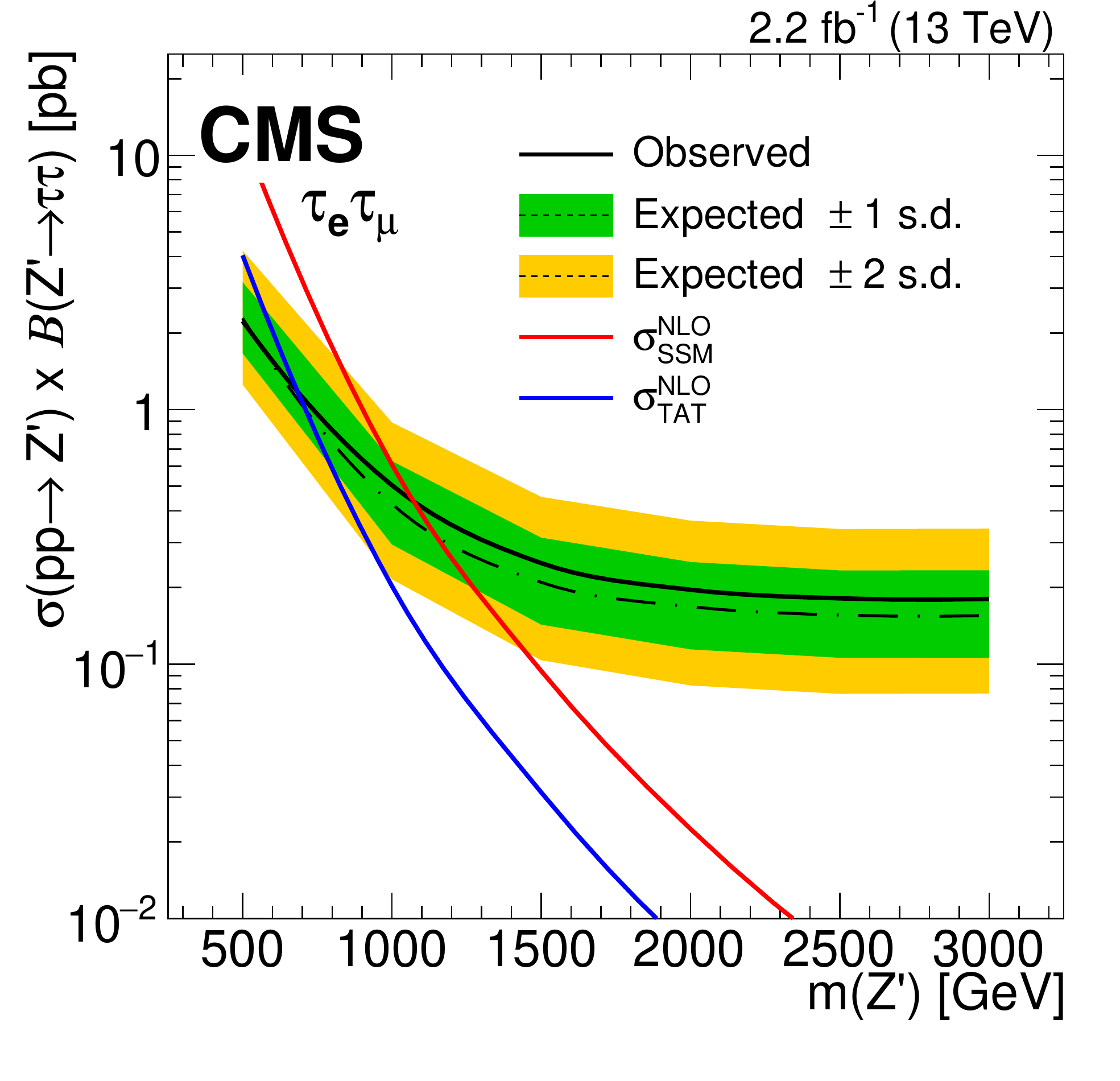}
    \includegraphics[width=0.42\textwidth]{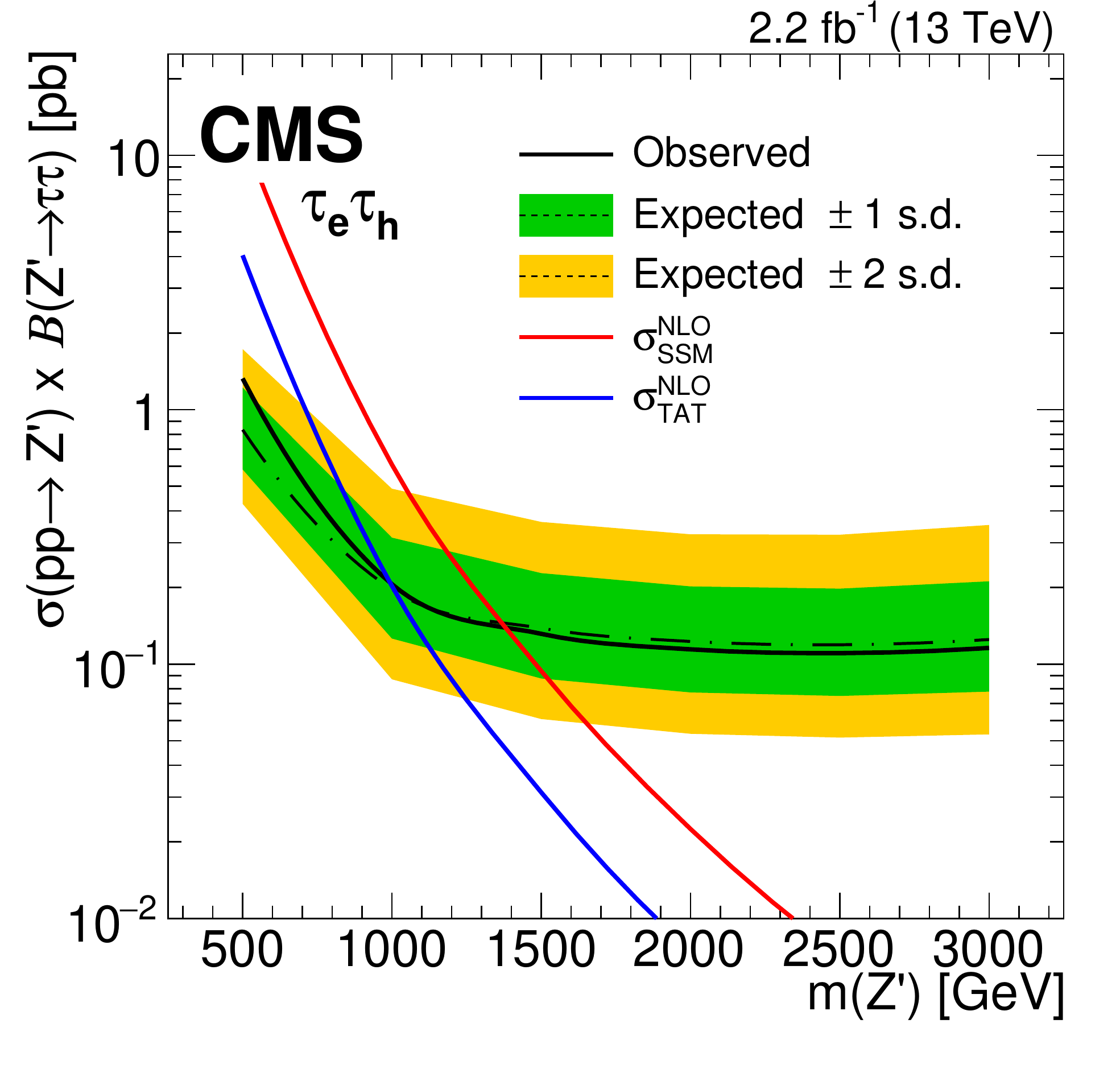}\\
    \includegraphics[width=0.42\textwidth]{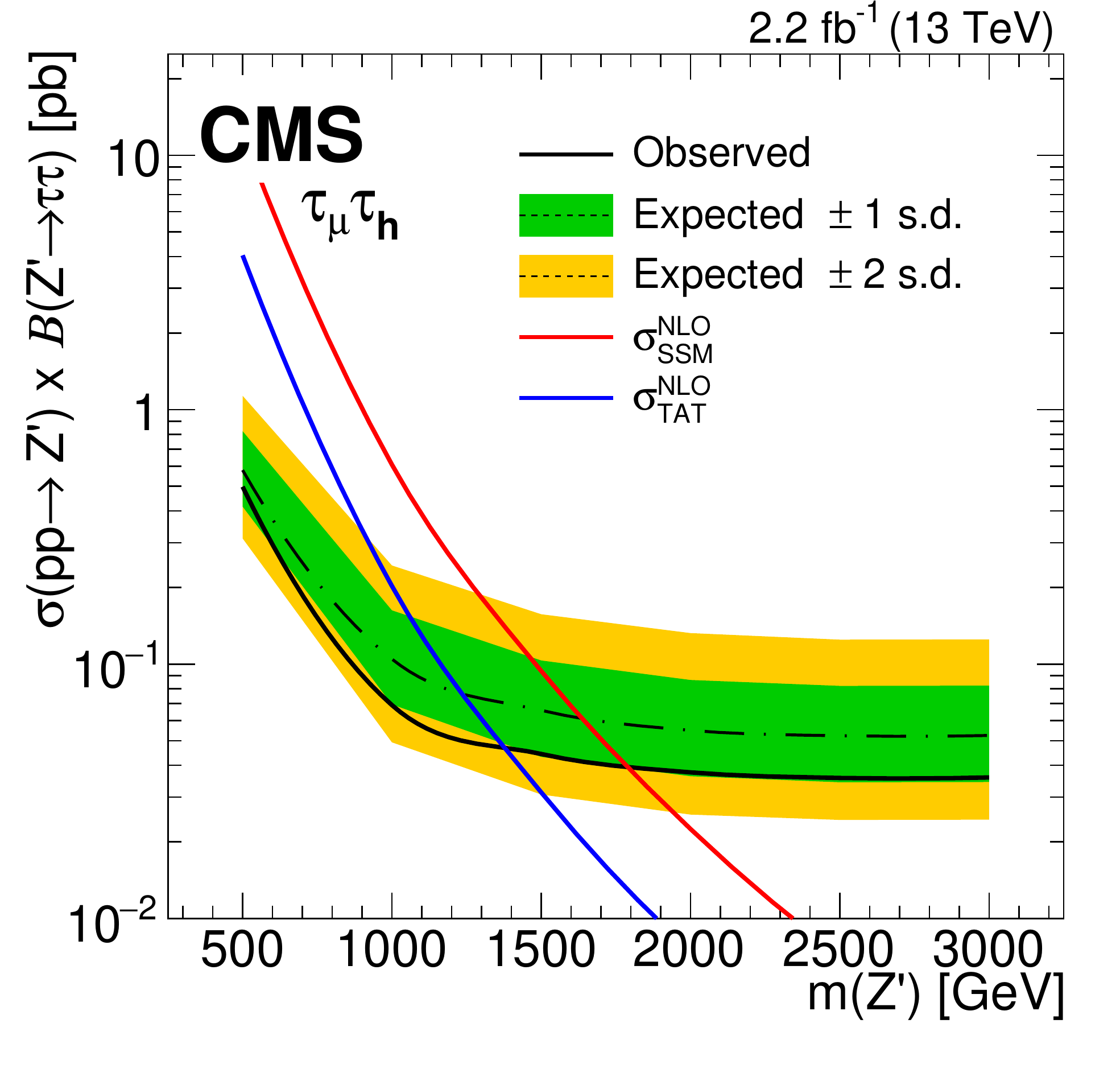}
    \includegraphics[width=0.42\textwidth]{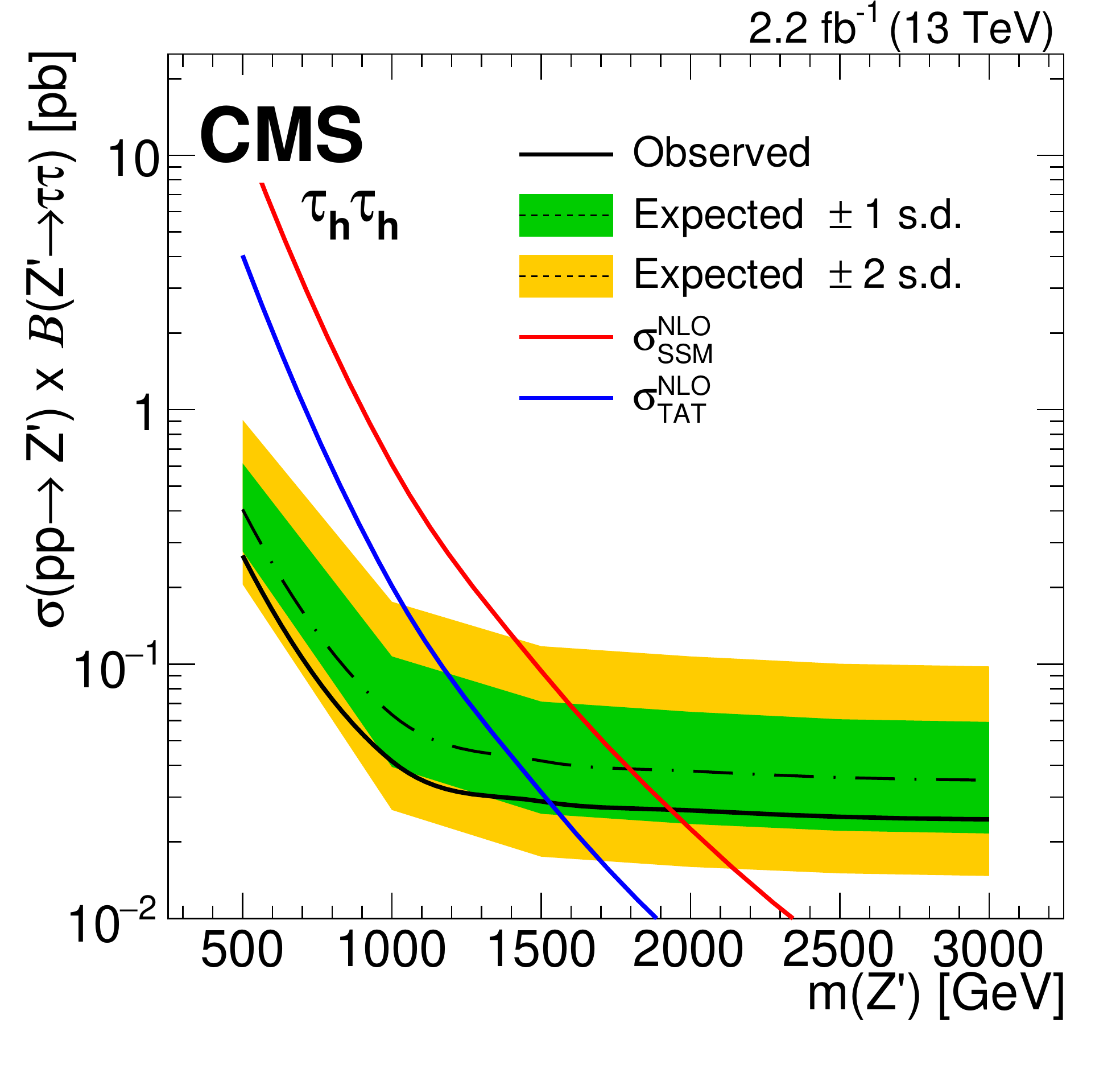}\\
    \includegraphics[width=0.42\textwidth]{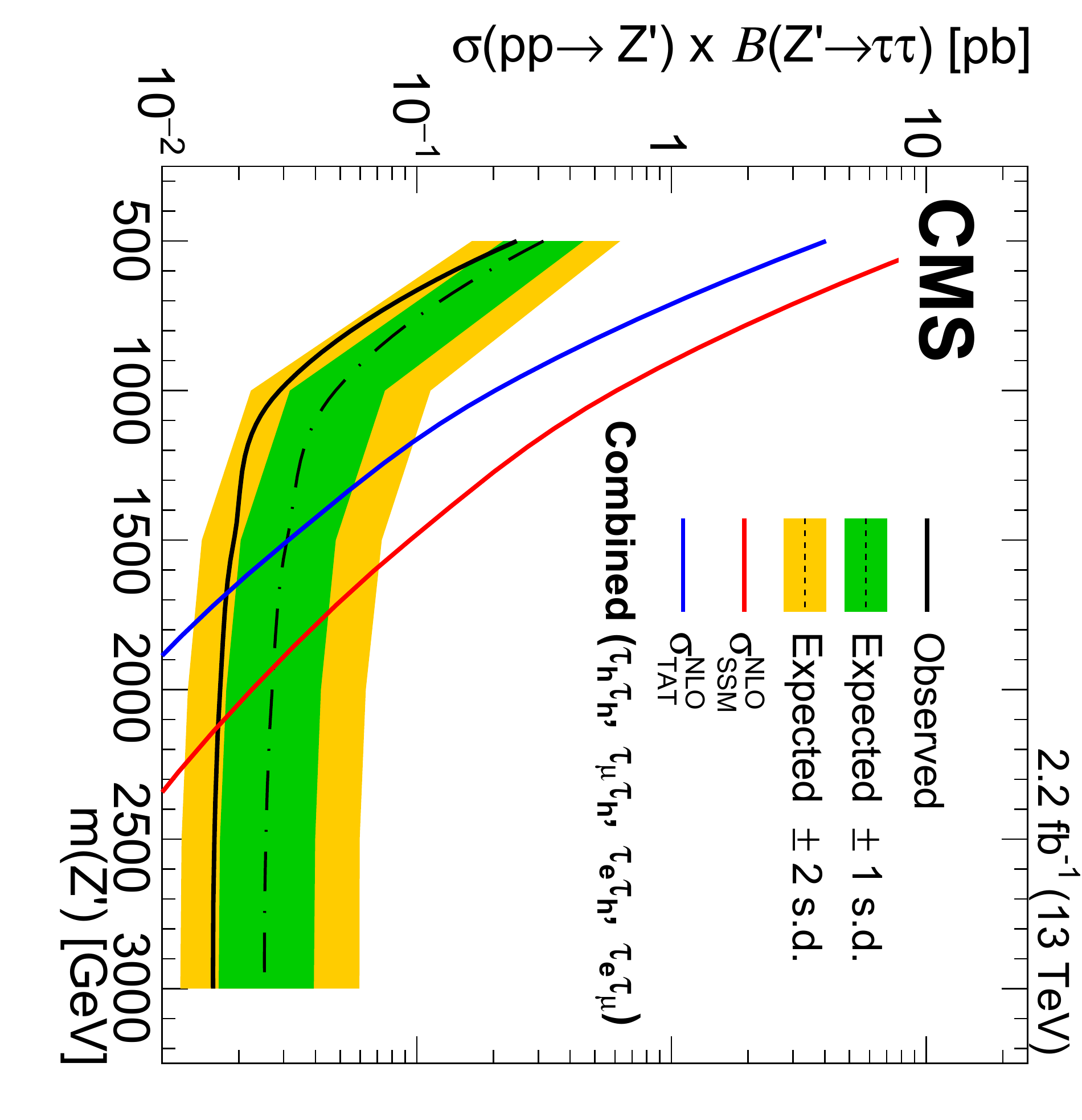}
    \caption{The observed 95\% CL upper limits on the product of the cross section and branching fraction into $\tau$ lepton pairs as a function of the \PZpr mass $m(\PZpr)$ (solid
black lines), for the (top left) \emu, (top right) \etau, (middle left) \mutau, and (middle right) \ditauhad final states, and (bottom) for the combination of the
four channels. The expected limits (dash-dotted lines) with one and two standard deviation (s.d.) uncertainty bands are also shown.  The predictions of the NLO
theory cross sections in the SSM and TAT models are represented by the red (lighter) and blue (darker) solid curves, respectively.}
 \label{fig:Limits}
\end{figure*}

The observed mass spectra shown in Fig.~\ref{fig:masses} do not reveal
evidence for new particles decaying to $\tau$ lepton pairs.  We
proceed to set upper limits on the product of the signal cross section and the
branching fraction for a \PZpr boson decaying to a $\tau$ lepton pair.
The modified frequentist construction
$\mathrm{CL_s}$~\cite{Junk,CLs,LHC-HCG} is used to determine these
upper limits at 95\% confidence level (CL) as a function of the \PZpr
mass for each $\tau\tau$ final state.  For each decay channel, and for
each bin of reconstructed mass above 85\GeV, the observed number of events is
fitted by a Poisson distribution whose mean is the sum of the total
SM expectation, determined as described in Section~\ref{Background
  estimation}, and a potential signal contribution determined from
simulation.  Systematic uncertainties are implemented as nuisance
parameters, which are profiled, and modeled with gamma or log-normal priors for
normalization parameters and Gaussian priors for shape uncertainties.

Figure~\ref{fig:Limits} shows the observed and expected limits on the product of the cross section and the branching fraction for the decays into $\tau$ lepton
pairs as functions of the \PZpr mass, together with the theoretical predictions from the SSM and TAT models.
The shaded bands represent one and two standard deviation
uncertainty intervals in the expected limits obtained using a large
sample of pseudo-experiments where the pseudodata are generated from
distributions corresponding to the background-only hypothesis.
The upper limit on the cross section
times branching fraction $\sigma(\Pp\Pp\to\PZpr)\,
\mathcal{B}(\PZpr\to\tau\tau)$ corresponds to the point where the
observed limit crosses the theory curve.
In the TAT model the gauge group structure is characterized by a reduced coupling to light fermions (1$^{\text{st}}$ and 2$^{\text{nd}}$ generations) and an enhanced coupling to heavy fermions
(3$^{\text{rd}}$ generation). Because light quark annihilation (e.g. $\PQu\PAQu \to \PZpr$) is the dominant \PZpr production mechanism at the LHC, the cross section $\sigma(\Pp\Pp\to\PZpr)$ in the
TAT model is suppressed (relative to that in the SSM model) because of the reduced coupling to light quarks. On the other hand, the branching fraction $\mathcal{B}(\PZpr\to\tau\tau)$ in the TAT model
is enhanced as a result of the stronger coupling to $\tau$ leptons. With the TAT parameters in Refs.~\cite{TopcolorAssistedTechnicolor1,TopcolorAssistedTechnicolor2}, the suppression in the cross
section outweighs the increase of the branching fraction. Overall, the product of the cross section and branching fraction $\sigma(\Pp\Pp\to\PZpr)\, \mathcal{B}(\PZpr\to\tau\tau)$ in the TAT model
is approximately one-third of the value in the SSM model.
Combining the four final states, we exclude \PZprrimeSSM and \PZprrimeTAT models with masses less than 2.1\TeV
(1.9\TeV expected) and 1.7\TeV\,(1.5\TeV expected), respectively, at 95\% CL.

\section{Summary}
\label{Summary}
A search for heavy resonances decaying to a tau lepton pair has
been performed by the CMS experiment, using a data sample of
proton-proton collisions at $\sqrt{s} = 13\TeV$ collected in 2015,
corresponding to an integrated luminosity of 2.2\fbinv.
The tau leptons are reconstructed in their decays to an electron (\taue) and muon
(\taumu), and in their hadronic decays (\tauh).  The observed invariant mass
spectra in the \mutau, \etau, \ditauhad, and \emu channels are
measured and are found to be consistent with expectations from the
standard model.  Upper limits at 95\% confidence level are derived for
the product of the cross section and branching fraction for a \PZpr
boson decaying to a tau lepton pair, as a function of the \PZpr mass.
The presence of \PZpr bosons decaying to a tau lepton pair is
excluded for \PZpr masses below 2.1\TeV in the sequential
standard model. This is the first search for heavy resonances decaying to a tau lepton pair
using events from proton-proton collisions at $\sqrt{s} = 13\TeV$, already extending previous limits~\cite{ZpCms7,ZpAtlas7, ZpAtlas8}
for this final state using the data sample collected in 2015.
In the topcolor-assisted technicolor model, which predicts \PZpr bosons that exhibit enhanced couplings to third-generation fermions,
the presence of \PZpr bosons decaying to a tau lepton pair is excluded for \PZpr masses below 1.7\TeV, resulting in the most stringent limit to date.

\begin{acknowledgments}
We congratulate our colleagues in the CERN accelerator departments for the excellent performance of the LHC and thank the technical and administrative staffs at CERN and at other CMS institutes for their contributions to the success of the CMS effort. In addition, we gratefully acknowledge the computing centers and personnel of the Worldwide LHC Computing Grid for delivering so effectively the computing infrastructure essential to our analyses. Finally, we acknowledge the enduring support for the construction and operation of the LHC and the CMS detector provided by the following funding agencies: BMWFW and FWF (Austria); FNRS and FWO (Belgium); CNPq, CAPES, FAPERJ, and FAPESP (Brazil); MES (Bulgaria); CERN; CAS, MoST, and NSFC (China); COLCIENCIAS (Colombia); MSES and CSF (Croatia); RPF (Cyprus); SENESCYT (Ecuador); MoER, ERC IUT, and ERDF (Estonia); Academy of Finland, MEC, and HIP (Finland); CEA and CNRS/IN2P3 (France); BMBF, DFG, and HGF (Germany); GSRT (Greece); OTKA and NIH (Hungary); DAE and DST (India); IPM (Iran); SFI (Ireland); INFN (Italy); MSIP and NRF (Republic of Korea); LAS (Lithuania); MOE and UM (Malaysia); BUAP, CINVESTAV, CONACYT, LNS, SEP, and UASLP-FAI (Mexico); MBIE (New Zealand); PAEC (Pakistan); MSHE and NSC (Poland); FCT (Portugal); JINR (Dubna); MON, RosAtom, RAS, and RFBR (Russia); MESTD (Serbia); SEIDI and CPAN (Spain); Swiss Funding Agencies (Switzerland); MST (Taipei); ThEPCenter, IPST, STAR, and NSTDA (Thailand); TUBITAK and TAEK (Turkey); NASU and SFFR (Ukraine); STFC (United Kingdom); DOE and NSF (USA).

\hyphenation{Rachada-pisek} Individuals have received support from the Marie-Curie program and the European Research Council and EPLANET (European Union); the Leventis Foundation; the A. P. Sloan Foundation; the Alexander von Humboldt Foundation; the Belgian Federal Science Policy Office; the Fonds pour la Formation \`a la Recherche dans l'Industrie et dans l'Agriculture (FRIA-Belgium); the Agentschap voor Innovatie door Wetenschap en Technologie (IWT-Belgium); the Ministry of Education, Youth and Sports (MEYS) of the Czech Republic; the Council of Science and Industrial Research, India; the HOMING PLUS program of the Foundation for Polish Science, cofinanced from European Union, Regional Development Fund, the Mobility Plus program of the Ministry of Science and Higher Education, the National Science Center (Poland), contracts Harmonia 2014/14/M/ST2/00428, Opus 2013/11/B/ST2/04202, 2014/13/B/ST2/02543 and 2014/15/B/ST2/03998, Sonata-bis 2012/07/E/ST2/01406; the Thalis and Aristeia programs cofinanced by EU-ESF and the Greek NSRF; the National Priorities Research Program by Qatar National Research Fund; the Programa Clar\'in-COFUND del Principado de Asturias; the Rachadapisek Sompot Fund for Postdoctoral Fellowship, Chulalongkorn University and the Chulalongkorn Academic into Its 2nd Century Project Advancement Project (Thailand); and the Welch Foundation, contract C-1845.
\end{acknowledgments}

\bibliography{auto_generated}
\cleardoublepage \appendix\section{The CMS Collaboration \label{app:collab}}\begin{sloppypar}\hyphenpenalty=5000\widowpenalty=500\clubpenalty=5000\textbf{Yerevan Physics Institute,  Yerevan,  Armenia}\\*[0pt]
V.~Khachatryan, A.M.~Sirunyan, A.~Tumasyan
\vskip\cmsinstskip
\textbf{Institut f\"{u}r Hochenergiephysik,  Wien,  Austria}\\*[0pt]
W.~Adam, E.~Asilar, T.~Bergauer, J.~Brandstetter, E.~Brondolin, M.~Dragicevic, J.~Er\"{o}, M.~Flechl, M.~Friedl, R.~Fr\"{u}hwirth\cmsAuthorMark{1}, V.M.~Ghete, C.~Hartl, N.~H\"{o}rmann, J.~Hrubec, M.~Jeitler\cmsAuthorMark{1}, A.~K\"{o}nig, I.~Kr\"{a}tschmer, D.~Liko, T.~Matsushita, I.~Mikulec, D.~Rabady, N.~Rad, B.~Rahbaran, H.~Rohringer, J.~Schieck\cmsAuthorMark{1}, J.~Strauss, W.~Waltenberger, C.-E.~Wulz\cmsAuthorMark{1}
% ...
\vskip\cmsinstskip
% ...
1:~~Also at Vienna University of Technology, Vienna, Austria\\
% ...

\end{sloppypar}
\end{document}